\documentclass[journal=jacsat,manuscript=article]{achemso}

\usepackage[version=3]{mhchem} 
\usepackage{xcolor}

\author{Pavel D. Terekhov}
\affiliation{Electrooptics and Photonics Engineering Department,  Ben-Gurion University, Beer-Sheva 8410501, Israel}
\alsoaffiliation{ITMO University, 49 Kronversky Ave., 197101, St. Petersburg, Russia}
\email{terekhovpd@gmail.com}
\author{Andrey B. Evlyukhin}
\affiliation{ITMO University, 49 Kronversky Ave., 197101, St. Petersburg, Russia}
\alsoaffiliation{Moscow Institute of Physics and Technology, 9 Institutsky Lane, Dolgoprudny 141700, Russia}
\email{a.b.evlyukhin@daad-alumni.de}
\author{Dmitrii Redka}
\affiliation{Saint Petersburg Electrotechnical University “LETI” (ETU), 5 Prof. Popova Street, St. Petersburg 197376, Russia}
\author{Valentin S. Volkov}
\affiliation{Moscow Institute of Physics and Technology, 9 Institutsky Lane, Dolgoprudny 141700, Russia}
\author{Alexander S. Shalin}
\affiliation{ITMO University, 49 Kronversky Ave., 197101, St. Petersburg, Russia}
\author{Alina Karabchevsky}
\affiliation{Electrooptics and Photonics Engineering Department, Ben-Gurion University, Beer-Sheva 8410501, Israel}
\email{alinak@bgu.ac.il}

\title[Magnetic Octupole Response of Dielectric Oligomers]
  {Magnetic Octupole Response of Dielectric Oligomers}


\begin{document}







\begin{abstract}
The development of new approaches to tuning the resonant magnetic response of simple all-dielectric nanostructures is very important in modern nanophotonics.
Here we show that a resonant magnetic octupole (MOCT) response can be obtained  by dividing a solid rectangular silicon block to an oligomer structure with the introduction of narrow gaps between four nanocubes.
We control and tune the spectral position of the MOCT resonance by varying the distance between the nanocubes.  
We demonstrate  that several magnetic hot-spots related to the MOCT resonance can be located in the gaps creating a strong magnetic field gradient in free space. 
We observe that the resonant excitation of the MOCT moment leads to a significant enhancement of light absorption in the system at the spectral region, where light absorption in bulk silicon is weak.
The results of this work can be applied to design new composite antennas and metamaterials based on complex building blocks, energy harvesting devices and molecular trapping with magnetic hot-spots.
\end{abstract}

\section{Introduction}
Dielectric nanophotonics is one of the most actively developing fields in photonics research. \cite{staude2017metamaterial,kivshar2018all,kamali2018review}
A variety of applications of dielectric nanostructures in technological devices has led to a growing interest of scientific groups over the globe.
One very important property of dielectric structures in comparison to their metal counterparts is the opportunity to control electric and magnetic components of light due to the excitation of electric and magnetic multipole resonances\cite{evlyukhin2010optical,evlyukhin2011multipole,evlyukhin2012demonstration}  with the simultaneous accumulation of electromagnetic energy. \cite{evlyukhin2012demonstration,zywietz2014laser,miroshnichenko2015nonradiating}
Additionally, dielectric high-index particles have commercial value due to the low resonant absorption in optical range, \cite{basharin2015dielectric} while plasmonic structures experience significant Ohmic losses. \cite{khurgin2015deal,karabchevsky2016tuning,karabchevsky2009theoretical}
To study  scattering of light by dielectric nanostructures one can use the multipole decomposition approach, already widespead in scientific investigations \cite{alaee2019exact, yang2017multimode,staude2013tailoring,terekhov2017resonant,chen2019multipolar,grahn2012electromagnetic,terekhov2019multipole,muhlig2011multipole,suzuki2019multipole,hancu2013multipolar,shamkhi2019transverse,luk2017hybrid,savinov2014toroidal,powell2017interference} including different spectral ranges \cite{terekhov2017destructive,balezin2018electromagnetic}.
State-of-the-art literature reports that the multipole responses can be tuned by changing particles' geometry \cite{terekhov2017multipolar}, size, aspect ratio, material dispersion and the refractive index of surrounding media.
Several recent studies have focused on high-order multipole excitations.\cite{zhu2018giant,liu2017beam,terekhov2019broadband,alaee2016phase,gurvitz2019high}

Owing to the large value of refractive index and low losses in near-infrared \cite{palik1997handbook} silicon is the most suitable material for the resonant dielectric nanophotonics \cite{evlyukhin2010optical} and structures development.  \cite{wang2018programming,deng2018sharp,terekhov2019enhanced,voroshilov2015light}
For instance, silicon nanostrips placed on optical waveguide allows for probing forbidden overtone transitions. \cite{katiyi2018si}
If placed on top of a lossy plasmonic material, silicon nanostrip allows for the realization of the cloaking effect and manipulation with waveguide's evanescent fields.\cite{galutin2017invisibility}

Dimers, oligomers and other dielectric nanostructures (fabricated from silicon) are used for magnetic field concentration and enhancement. \cite{baryshnikova2017magnetic,bakker2015magnetic}
In this work we study the optical properties of the silicon oligomer which supports the resonant excitation of a magnetic octupole moment and allows it to be controlled using structure parameters.
We show that a solid block of crystalline silicon does not support magnetic octupole resonances, and that simply cutting it enables a resonant magnetic octupole response of the resulting silicon oligomer.

This effect leads to both the magnetic field enhancement inside the structure's slits and to increased light absorption by the structure.
Moreover, the location of magnetic hotspots can be switched using incident light polarization.
The suggested structures can be used to design modern optical devices and for efficient light control using magnetic octupole excitation.
It is worth noting, that the achieved magnetic octupole response appears in an unusual part of the spectrum: for a bulk structure of comparable size, it would appear at far shorter wavelengths.
In general, tailoring the resonant response of high-order multipole moments over the optical spectral range opens new opportunities in practical applications, e.g. sensors, detectors, and selective or directive nanoantennas.

\begin{figure}[tb]
  \centering
  \includegraphics[width=\linewidth]{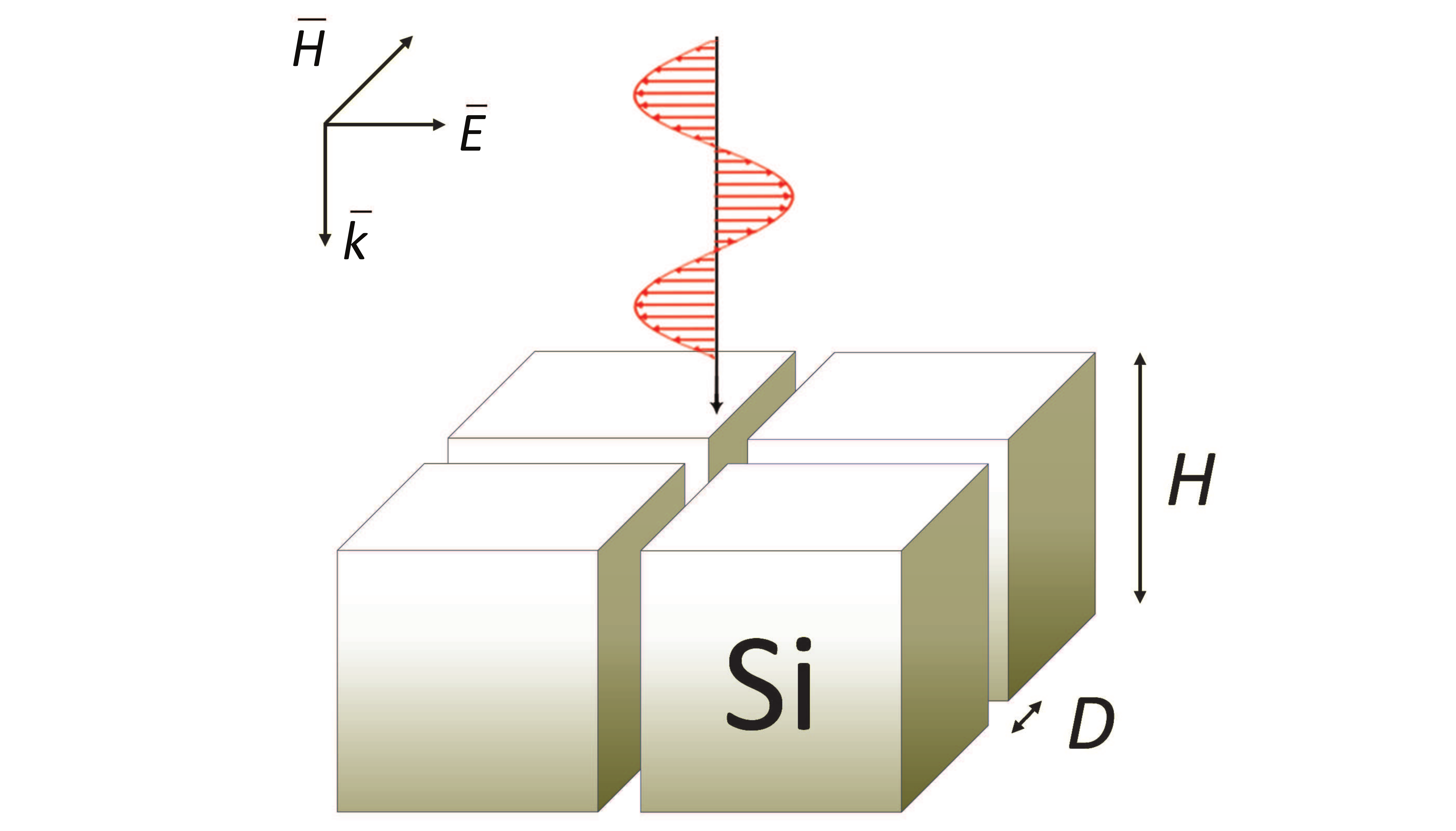}
  \caption{The artistic representation of the oligomer of silicon cubes.}
  \label{fig:figure1}
\end{figure}

\section{Theoretical Background} 
\label{sec:theoretical_background}

Here we use multipole decomposition approach described in \cite{evlyukhin2016optical,alaee2018electromagnetic,alaee2019exact,evlyukhin2019multipole} including  multipole moments up to the magnetic octupole  (MOCT) term.
To study the multipole contributions to a scattering cross-section of full oligomer structure presented in Fig. \ref{fig:figure1} we consider it as an unite system. 
The center of mass of the system is in the origin of the coordinate system.

In our approximation a scattering cross-section of a particle in a homogeneous host medium can be presented as (see  \cite{evlyukhin2019multipole} for details):
\begin{eqnarray}\label{SigmaSca}
C_{\rm sca}&\simeq&\frac{k_0^4}{6\pi\varepsilon_0^2 |\textbf{E}_{inc}|^2}|\mathbf{D}|^2+\frac{k_0^4\varepsilon_d\mu_0}{6\pi\varepsilon_0|\textbf{E}_{inc}|^2}|{\bf m}|^2\nonumber\\
&&+\frac{k_0^6 \varepsilon_d}{720\pi\varepsilon_0^2
|\textbf{E}_{inc}|^2}|\hat{Q}|^2+\frac{k_0^6
\varepsilon_d^2\mu_0}{80\pi\varepsilon_0 |\textbf{E}_{inc}|^2} |\hat{M}|^2\\
&&+\frac{k_0^8 \varepsilon_d^2}{1890\pi\varepsilon_0^2
|\textbf{E}_{inc}|^2}|\hat{O}|^2 + \frac{k_0^8 \varepsilon_d^3 \mu_0}{1890 \pi k_0^8 \varepsilon_0|\textbf{E}_{inc}|^2}|\hat{O}_m|^2 \nonumber,
\end{eqnarray} 
where $\textbf{E}_{inc}$ is the electric field amplitude of the incident light wave, $\varepsilon_d=n_{d}^{2}$ is the dielectric permittivity of the surrounding medium, \textbf{\( \varepsilon _{0}\) } is the vacuum electric permittivity and \( v_{d}=c/\sqrt{\varepsilon_{d}} \)  is the light speed in the surrounding medium; \( k_{0} \) and \(k_{d}\) are the wavenumbers in vacuum and in the surrounding medium, correspondingly. $\textbf{m}$ is the magnetic dipole moment (MD) of a particle; $\mathbf{D}$ is the total electric dipole moment (TED); $\hat{Q}$, $\hat{M}$, $\hat{O}$ and $\hat{O}_m$ are the electric quadrupole tensor (EQ), the magnetic quadrupole tensor (MQ), the tensor of electric octupole (OCT), and the tensor of magnetic octupole (MOCT), respectively. 
Here $||$ denotes the sum of squared tensor components.
Note that these tensors are symmetric and traceless, and in tensor notation, e.g., $\hat{Q}$ is equal to $Q_{\alpha\beta}$ (for quadrupole moments) and $\hat{O}$ is equal to $O_{\alpha\beta\gamma}$ (for octupole moments), where subscript indices denote components (e.g, $\alpha={x,y,z}$). \cite{evlyukhin2016optical}
Let us also show the expressions used here for the Cartesian electric and magnetic octupole moments, expanding the multipole decomposition at \cite{alaee2018electromagnetic} :
\begin{equation}\label{O}
  \hat{O} = \frac{15 i}{\omega} \int_{V_s} \frac{j_2 (k_d r')}{(k_d r')^2} \left(\mathbf{j r' r'} + \mathbf{r' j r'} + \mathbf{r' r' j} - \hat{A}\right) d \mathbf{r'},
\end{equation}
\begin{eqnarray}\label{Om}
  \hat{O}_m& =& \frac{105}{4} \int_{V_s} \frac{j_3 (k_d r')}{(k_d r')^3} \left(  [\mathbf{r'} \times \mathbf{j}]\mathbf{r' r'} + \mathbf{r'} [\mathbf{r'} \times \mathbf{j}] \mathbf{r'}\right.\nonumber\\ 
  & &+\left. \mathbf{r' r'} [\mathbf{r'} \times \mathbf{j}] - \hat {A'}\right) d \mathbf{r'},
\end{eqnarray}
where vector $\mathbf{j}$ is the electric current density induced in the scatterer by an incident light wave and $\mathbf{r'}$ is the radius vector of a volume element inside the scatterer; $j_2, j_3$ are the spherical Bessel functions, $V_s$ is a scatterer volume, and the tensors $\hat {A}$ and $\hat {A'}$ are
\begin{equation}
A_{\beta \gamma \tau} = \delta _{\beta \gamma} V_\tau + \delta_{\beta \tau} V_\gamma + \delta_{\gamma \tau} V_\beta,
\end{equation}
\begin{equation}
  A'_{\beta \gamma \tau} = \delta _{\beta \gamma} V'_\tau + \delta_{\beta \tau} V'_\gamma + \delta_{\gamma \tau} V'_\beta,
\end{equation}
where $\beta = x,y,z$; $\gamma = x, y, z$; $\tau = x, y, z$; $\delta _{\beta \gamma}$ is the Kronecker delta,
\begin{equation}
\mathbf{V} = \frac{1}{5} [2({\bf r}'\cdot{\bf j}){\bf r}'+ r'^2 {\bf j}],
\end{equation}
\begin{equation}
  \mathbf{V'} = \frac{1}{5} r'^2 \left [\mathbf{r'} \times \mathbf{j} \right].
\end{equation}
The combinations of three vectors (like $\mathbf{j r' r'}$) in (\ref{O}) and (\ref{Om}) are the tensor products of the  corresponding vectors. Detailed derivations are presented in Ref.  \cite{evlyukhin2019multipole}.

The total scattering cross-section is obtained through the integration of the Pointing vector over a closed surface in the far-field zone and the normalization to the incident field intensity.

\begin{figure}[tb]
  \centering
  \includegraphics[width=1\linewidth]{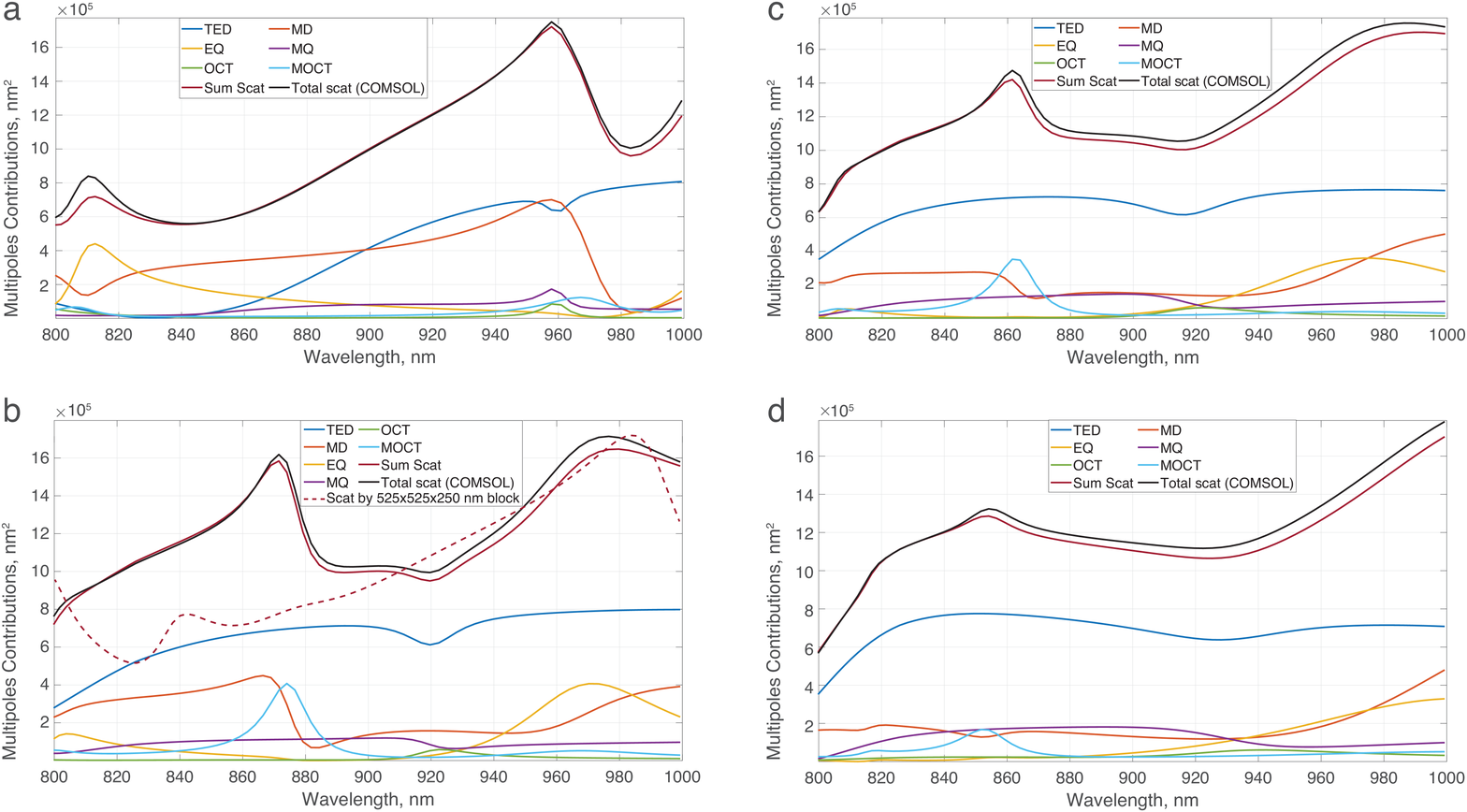}
  \caption{Scattering cross-section spectra and corresponding multipoles' contributions calculated for: (a) the single silicon block of height $H$ = 250 nm and base edge 500 nm; (b,c,d) the oligomer of silicon cubes. The distance between cubes in the oligomer is (b) $D$ = 25 nm (c) $D$ = 50 nm (d) $D$ = 100 nm. 'Sum Scat' states for the scattering cross-section as the sum of the multipole contributions; 'Total scat (COMSOL)' states for the total scattering cross-sections calculated directly in COMSOL.}
  \label{fig:figure2}
\end{figure}

\section{Results and discussion} 
\label{sec:results_and_discussion}

In this work, we study previously unrevealed MOCT-induced optical properties of silicon oligomers.
We learn how to use controllable resonant MOCT excitation to tailor magnetic hot-spots and resonant energy absorption.
The considered systems are the silicon block with dimensions equal to $500\times500\times250$ nm, which represents zero distance between silicon cubes in Fig. \ref{fig:figure1} and silicon oligomers composed of four Si cubes ($250\times250\times250$ nm) with distance between them $D$ = 25, 50 and 100 nm.
These structures are illuminated with a linearly polarized plane wave as shown in Fig. \ref{fig:figure1}.
To investigate scattering cross-section spectral resonances, we apply the multipole decomposition technique which shows good performance in all cases considered (Fig. \ref{fig:figure2}).
The almost perfect agreement between the sum of the multipole contributions and the directly calculated scattered cross-section proves that the multipole approach is sufficiently accurate.
Fig. \ref{fig:figure2} (a) shows the scattering cross-section spectrum calculated for the solid silicon block. 
One can note that both the resonant multipole contributions and the total scattering cross-section in this case significantly differ from the spectra in Fig. \ref{fig:figure2} (b,c,d) due to the introduction of inhomogeneity to the system.

In this way, the conversion of the solid block to the oligomer structure leads to a strong reconfiguration of electric and magnetic fields in the system and to a higher order multipole excitation. 
Surprisingly, the presence of narrow air gaps in the oligomer leads to the excitation of the MOCT moment in the considered spectral range.
Note that this is not related to the increase in the total structure size, since our calculations for solid blocks with the edge of 525 nm do not show MOCT resonances in the considered spectral range.
To visualize this, we show the total scattering cross-section of $525\times525\times250$ nm silicon block (dashed line in Fig. \ref{fig:figure2} (b)) in order to compare it with the case of the corresponding oligomer.
It can be seen that there is no resonant response between $\lambda = 850$ nm and $\lambda = 900$ nm; small resonant peak at $\lambda$ = 840 nm corresponds to EQ.

\begin{figure}[tb]
  \centering
  \includegraphics[width=\linewidth]{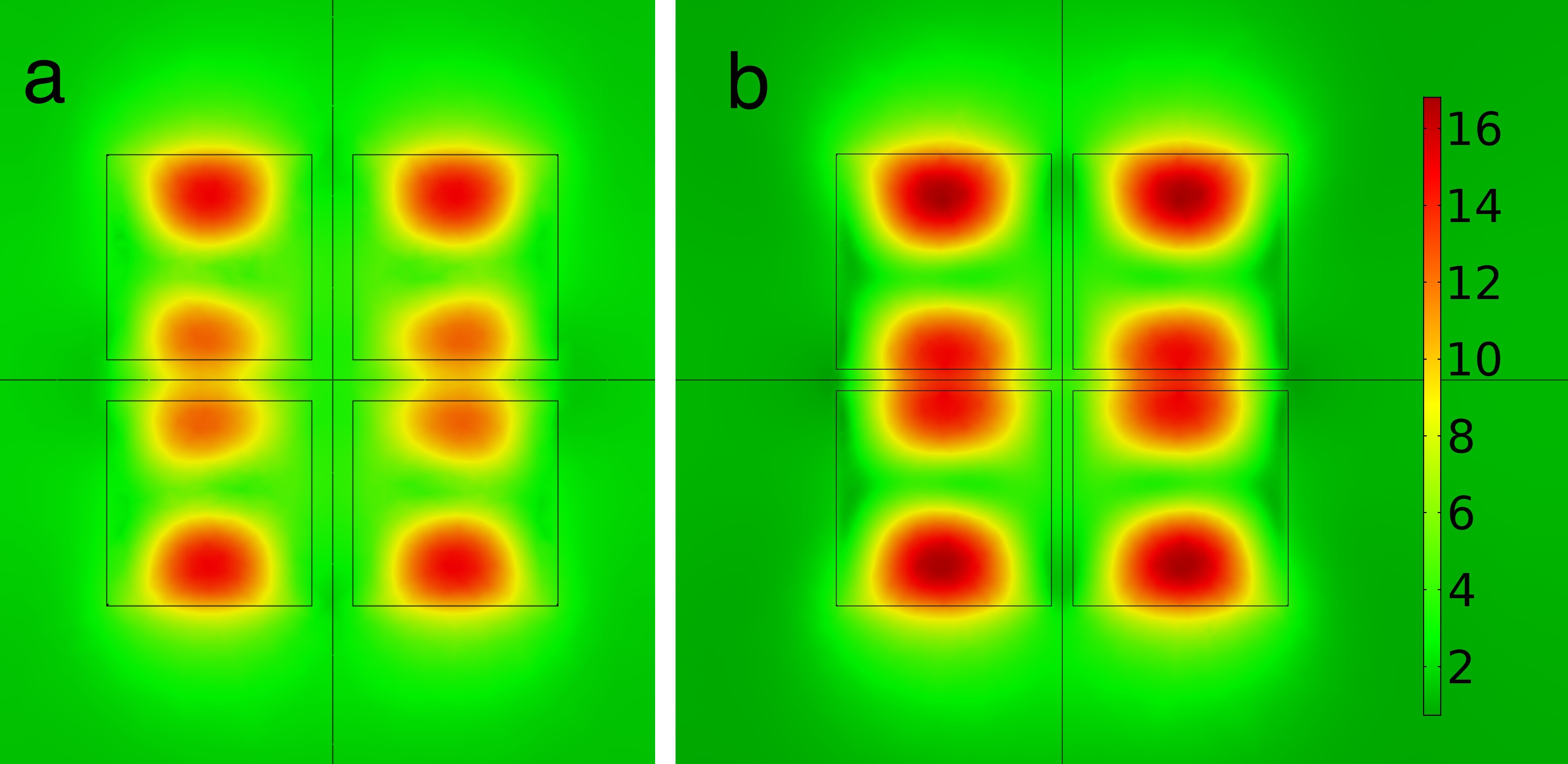}
  \caption{(a) Coefficient of the magnetic field amplitude enhancement in (xy) - plane (z = 0) of the silicon oligomer with (a) $D$ = 50 nm, $\lambda$ = 874 nm (b) $D$ = 25 nm, $\lambda = 863 nm$. Color bar is the same for both pictures.}
  \label{fig:figure3}
\end{figure}

Let us consider in detail the resonant excitation of the magnetic octupole moment at the wavelengths of 850 - 900 nm.
Cutting the solid block enables MOCT resonance weakening with increasting the intercube distance.
The resonant  MOCT peak occurs at $\lambda$ = 874 nm in Fig. \ref{fig:figure2} (b), at $\lambda$ = 863 nm in Fig. \ref{fig:figure2} (c) and at $\lambda$ = 852 nm in Fig. \ref{fig:figure2} (d).
While attenuating, the resonant peak also experiences a blue shift.
Moreover, the multipole decomposition and scattering cross-section do not depend on rotation of the incident plane wave polarization by 45$^\circ$ (not demonstrated in figures).

\begin{figure*}[tb]
  \centering
  \includegraphics[width=\linewidth]{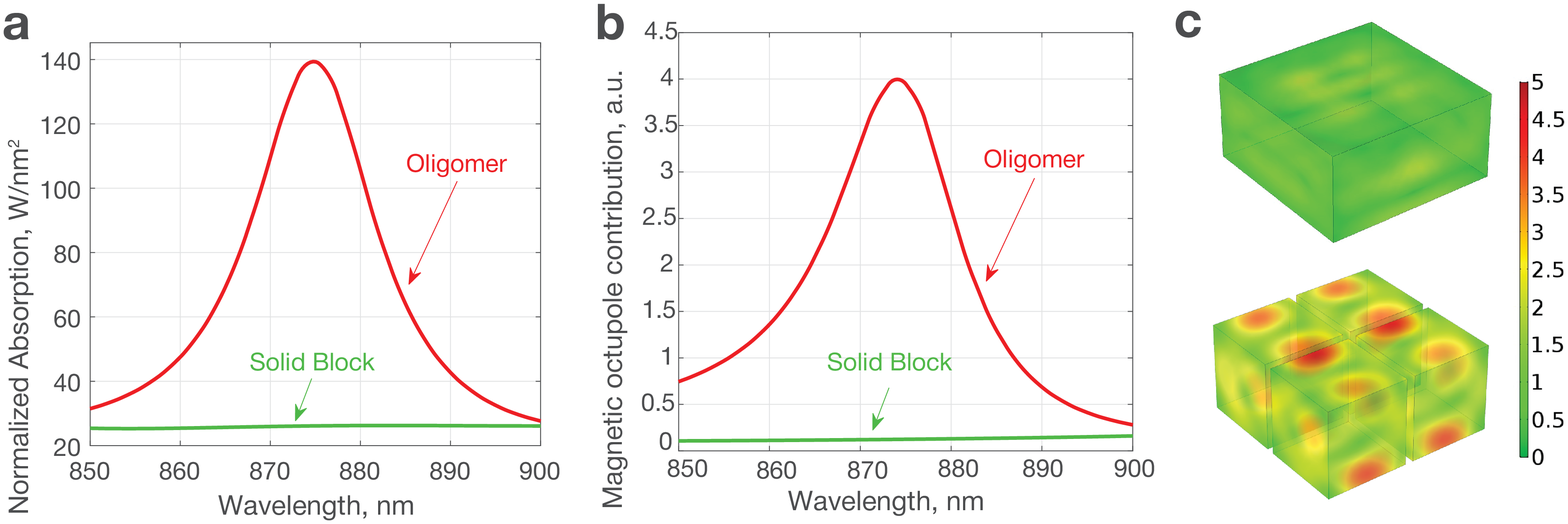}
  \caption{Spectra of the (a) normalized absorption power and (b) MOCT contribution to the scattering cross-section calculated for a single silicon block of height $H$ = 250 nm and base edge 500 nm (green lines) and the oligomer of silicon cubes with the distance between cubes $D$ = 25 nm (red lines). The absorption peak in the structure clearly corresponds to the resonant excitation of MOCT moment. (c) Normalized electric field inside the solid block (top) and oligomer (bottom) at $\lambda$ = 874 nm. One can see that resonant MOCT response provides strong electric field concentration leading to the resonant absorption in the silicon oligomer.}
  \label{fig:figure4}
\end{figure*}

Figure \ref{fig:figure3} shows the magnetic field magnitude at the MOCT resonance.
It is important for practical applications to be able to create the so-called magnetic hot spots in free space. \cite{baryshnikova2017magnetic}
Due to structuring, the total magnetic field in the gaps can be enhanced (comparing to the incident one) up to $\approx$ 10 times for $D$ = 50 nm (Fig. \ref{fig:figure3} (a)) and $\approx$ 14 times for $D$ = 25 nm (Fig. \ref{fig:figure3} (b)).
Strong local magnetic fields can be used to control or detect small quantum objects (quantum dots, atoms, and molecules) supporting magnetic optical transitions. \cite{degen2017quantum}
Magnetic hot-spots are also useful for spectroscopy \cite{sadrara2019electric}, better enhancement of Raman scattering, fluorescence and circular dichroism of molecules \cite{naeimi2018magnetic}, sensing \cite{yong2016narrow} and other applications.

The location of obtained hotspots can also be tuned using the incident light polarization.
The hotspots move to another air gaps if the polarization is rotated by 90$^\circ$ (i.e. if electric field polarized along perpendicular axis.
It is possible to exploit this effect to design magnetic switchers at the nanoscale.

In addition to the magnetic field enhancement, MOCT resonance can provide a strong electromagnetic absorption in the oligomer.
Figure \ref{fig:figure4} (a) shows the comparison of absorbed power for $\lambda$ between 850 nm and 900 nm for the solid silicon block and the oligomer with the distance $D$ = 25 nm between the cubes. 
Figure \ref{fig:figure4} (b) proves that the discovered energy absorption peak spectrally corresponds to the MOCT resonance.
It is worth noting that in this spectral range natural light absorption by silicon is small.
Therefor, air gaps in the oligomer structure cause strong absorption in silicon, despite its very small $Im(n) \approx 0.08$.
Fig. \ref{fig:figure4} (c) compares the electric field inside the silicon block and oligomer structure.
Clearly, resonant magnetic octupole response leads to a strong electric field concentration and, therefore, to the resonant absorption in the silicon oligomer.
The spectral position of the MOCT resonance and, hence, the position of the absorption peak can be changed by varying the distance between the cubes.
Such the tunable absorption can be widely used to control the energy concentration by dielectric structures and to design modern optical devices.

\section{Conclusion} 
\label{sec:conclusion}

In this work we study resonant MOCT excitation which leads to controlled magnetic hot-spots and resonant absorption by the nanostructure.
We analyze the multipole contributions to the scattering cross-section and reveal the excitation of magnetic octupole due to the coupling effects in this structure as compared to the single cube.
To the best of our knowledge, the resonant excitation of a magnetic octupole moment in dielectric oligomers demonstrated and analyzed in detail for the first time.
In this work, we show how to control MOCT resonant excitation and its spectral position.
In addition, we reveal its possible application to obtain magnetic hot-spots and to absorb electromagnetic energy in the nanostructure.
The magnetic field in the air gap between the nanocubes can be 14 times larger in comparison with incident one, and is even stronger inside the nanostructure.
Moreover, the magnetic hotspots can be switched between different slits simply by changing incident light polarization.
Use of oligomers as building blocks for metasurfaces promises even higher magnetic field enhancement due to potential excitation of so-called trapped modes.
Besides, resonant magnetic octupole response can be widely used for spectroscopy, sensing, small quantum objects detection, and many other promising applications.


\begin{acknowledgement}
Authors thank Svetlana Korinfskaya for the help with the artistic work.
This work has been supported by the Israel Innovation Authority-Kamin Program, Grant. No. 62045 (Year 2).
A.S.S acknowledges the support of the Russian Fund for Basic Research within the projects 18-02-00414, 18-52-00005. 
The development of the analytical approach  has been partially supported by the Russian Science Foundation Grant No. 18-79-10208. 
Support from the the Russian Fund for Basic Research within the projects 18-29-02089 is acknowledged as well.
Contribution of P.D.T. in the research was performed  as a part of the joint Ph.D. program between the BGU and ITMO.
\end{acknowledgement}

\bibliography{oligomer}

\end{document}